\documentstyle[11pt,a4]{article}
\newcommand{\V}{{\bf V}}
\newcommand{\W}{{\bf W}}
\newcommand{\Na}{{\bf N}}
\newcommand{\si}{\sigma}
\newcommand{\pr}{\prime}

\newcommand{\be}{\begin{equation}}
\newcommand{\ee}{\end{equation}}
\newcommand{\bea}{\begin{eqnarray}}
\newcommand{\ena}{\end{eqnarray}}
\newcommand{\beas}{\begin{eqnarray*}}
\newcommand{\enas}{\end{eqnarray*}}

\newcommand{\hg}{U_q\hat{g}}
\newcommand{\g}{U_qg}

\newcommand{\D}{\Delta}
\newcommand{\Sp}{{\bf S}}
\newcommand{\ot}{\otimes}
\newcommand{\up}{\uparrow}
\newcommand{\dow}{\downarrow}

\newcommand{\nn}{\nonumber}
\newcommand{\la}{\lambda}

\newcommand{\hot}{\hat{\otimes}}

\newcommand{\op}{\oplus}
\renewcommand{\H}{{\cal H}}

\newcommand{\Pro}{{\cal Q}}
\newcommand{\Proj}{{\cal P}}
\newcommand{\ab}[1]{{\rm{\bf #1}}}
\renewcommand{\dim}{{\rm dim}}
\newcommand{\id}{{\rm id}}
\newcommand{\gr}[2]{ {( #1,#2)} }
\newcommand{\re}[1]{(\ref{#1})}

\begin{document}
\hfill NBI-HE-98-04

\vspace{24pt}

\begin{center}
{ \large \bf
 Bethe Ansatz and Thermodynamic Limit of Affine \\
Quantum Group Invariant Extensions of the t-J Model}\footnote{This 
work was supported in part by INTAS grant 96-524}

\vspace{36pt}

J.Ambj\o rn\footnote{e-mail:{\sl ambjorn@nbivms.nbi.dk}}

\vspace{6pt}
{\sl Niels Bohr Institute}

{\sl Blegdamsvej 17, Copenhagen, Denmark}

\vspace{24pt}

A.Avakyan\footnote{e-mail:{\sl avakyan@lx2.yerphi.am}},
            T.Hakobyan\footnote{e-mail:{\sl hakob@lx2.yerphi.am}}

\vspace{6pt}
{\sl Yerevan Physics Institute,}

{\sl Br.Alikhanian st.2, 375036, Yerevan, Armenia}

\vspace{24pt}

A.Sedrakyan\footnote{e-mail:{\sl sedrak@nbivms.nbi.dk}; permanent
address: Yerevan Physics Institute},

\vspace{12pt}

{\sl Niels Bohr Institute}

{\sl Blegdamsvej 17, Copenhagen,  Denmark}

\vspace{36pt}
January 1998



\vfill
{\bf Abstract}
\end{center}

We have constructed a one dimensional exactly solvable model, which is based 
on the t-J model of strongly
correlated electrons, but which has additional quantum group symmetry, 
ensuring
the degeneration of  states. We use Bethe Ansatz technique to investigate
this model. The thermodynamic limit of the model is considered and
equations for different density functions written down.
These  equations  demonstrate that
the additional colour degrees of freedom of the model
behave as in a gauge theory, namely an arbitrary distribution of colour indices
over particles leave invariant the  energy of the ground state and the  
excitations. The $S$-matrix of the model is shown to be the product
of the ordinary $t-J$ model $S$-matrix and the unity matrix in the
colour space.

\vfill


\newpage

\section{Introduction}
Since the discovery of high-T$_c$ cuprate superconductivity the one
dimensional physics of strongly correlated electrons has been  in focus
in many publications \cite{KE}. The Hubbard \cite{H} and $t-J$ models 
\cite{ZhR}
are such examples , motivated in part by the fact that high-T$_c$
compounds display antiferromagnetism in the absence of doping. The $t-J$ model
was proposed by Zhang and Rice \cite{ZhR} and describes strongly correlated
electrons with anntiferromagnetic exchange interaction.

The interest in one dimensional physics grew after Andersons claim
\cite{An} that two-dimensional systems may have features in
common with one dimensional
systems. In addition it should be mentioned that powerful methods in 1D such
as bosonization, 2D conformal field theory and in particular the Bethe ansatz
technique allow the detailed study of such systems.

The $t-J$ model may be used as well for heavy fermion system \cite{S1,S2}.

At the supersymmetric point $J= 2t$ the $t-J$ model becomes exactly integrable
\cite{S1, SU,L,B,SA,EK,F} because the Hamiltonian can be represented as a
graded permutation in a superalgebra of two fermions and one boson.

In \cite{HS,HS2,AAH} we developed the technique for construction of a family of
spin chain Hamiltonians and their fermionic representations, which have the 
same energy levels as some basic model($ XXZ$, Hubbard, $t-J$ or others) 
but with
huge degeneracy as a result of an affine quantum symmetry added to the
basic model. We called this procedure an affinization of the model.

The first example of this type of model was constructed in \cite{RA}, giving
rise to the Hubbard Hamiltonian in the infinite repulsion limit.

In \cite{HS2} we have fermionized the  simplest examples of this
newly defined family of models and have shown that it leads to extensions
of one-band Hubbard Hamiltonians. The $\eta$-pairing mechanism introduced
by Yang  \cite{Y,YZ} was found in one of examples in addition to other
exactly solvable Hubbard models with superconducting ground state 
\cite{St,Ov,BKS,MC,Schad}. The essential property of this extensions is 
the fact that,
besides the ordinary electron hopping and Hubbard interaction terms, they
contain also bond-charge interaction, pair-hopping and nearest-neighbour 
interaction terms. In \cite{AAH} the $SU(N)$ affinization of $t-J$ model
was carried out, giving rise a model where the spin-spin coupling term
consists of interaction between the total spins(i.e. the sum of the spins 
of all band) at nearest-neighbour sites. The presence of the affine symmetry, 
which
ensures the degeneracy of levels exponentially proportional to the length
(area) of the space, might lead to a new type of string theory.

In this article we define an extension of $t-J$ model such that an affine
quantum group symmetry is present, and we use the Bethe 
ansatz technique  to solve
the model. We find the S-matrix excitations on empty background, the ground
state and construct the thermodynamic limit of the model. As one might
expect, the S-matrix of the excitations on empty background consists of the  
S-matrix of the ordinary $t-J$ model multiplied by the 
unity matrix in the additional
space of ``colours''. Therefore the Bethe equations are not different from
ones for t-J model, but  the rapidities presented in 
equations correspond to particles with the arbitrary colours. The degeneracy
of the corresponding n-particle states come from arbitrary partitions of
the colour indices over particles. The same is true for the ground state.
The situation is exactly  as in gauge theories  if we distinguish
the states which differs by  pure gauge transformations. 
All this results are presented in the Section 5 of the article.

The thermodynamic limit of the model with corresponding equations
are represented in the Section 6, where we also shown that the S-matrix of 
our model in an arbitrary background is equal to ordinary $t-J$ model
S-matrix multiplied by the Kronecker symbols over the additional colour
indices.    

\setcounter{equation}{0}
\section{Quantum group invariant Hamiltonians for reducible
        representations}

Let $\V=\oplus_{i=1}^N \V_{\la_i}$ be a direct sum of finite dimensional
irreducible representations $\V_{\la_i}$ of quantum group $\g$ 
\cite{Dr86,J85, J86_1}.
We denote by $\V(x_1,\dots,x_N)$ the
representation with spectral parameters $x_i$
of the corresponding affine quantum group $\hg$ \cite{J86_1}:
\be
\label{Vdecompos}
 \V(x_1,\dots,x_N)=\bigoplus_{i=1}^M \Na_{\la_i}\hot \V_{\la_i}(x_i),
\ee
where all the $\V_{\la_i}(x_i)$ are $M$ nonequivalent irreps and
$\Na_{\la_i}\simeq \ab{C}^{N_i}$ have dimensions equal to
the multiplicity of $\V_{\la_i}(x_i)$
in $\V(x_1,\dots,x_N)$.
Note that $\sum_{i=1}^M N_i=N$.
The $\hat{~}$ over the tensor product signifies that $\hg$ does not act on
$\Na_{\la_i}\hot \V_{\la_i}(x_i)$ by means of co-multiplication
$\D$ but instead acts as $\id\ot g$.

In \cite{HS2} the general matrix form of the intertwining operator
\bea
\label{int}
 && H(x_1,\dots,x_N):\nn\\
&& \quad \V(x_1,\dots,x_N)\ot \V(x_1,\dots,x_N)\rightarrow
\V(x_1,\dots,x_N)\ot \V(x_1,\dots,x_N),
\\
 && \quad {[}H(x_1,\dots,x_N),\D(a){]}=0, \qquad \forall
 a\in\hg\nn
\ena
had been
written using the projection operators
\be
\label{proj}
X^a_b=|a\rangle\langle b|
\ee
Here the  vectors
$|a\rangle$ span the space $\V$. In accordance with the decomposition
\re{Vdecompos} we will
use the double index $a=\gr{n_i}{a_i}$, $i=1,\dots,M$
where the first
index  $n_i=1,\dots,N_i$ characterises the multiplicity of
$\V_{\la_i}$ and the second one $a_i=1,\dots,\dim{\V_{\la_i}}$
is the vector index of $\V_{\la_i}$.
Then the intertwining operator \re{int}   is
\bea
\label{HactMat}
H(A,B)=
\sum_{i,j=1}^M
\left(
\sum_{n_i,n_j,m_i,m_j}
A_{ij}\mbox{}_{n_in_j}^{m_im_j}
\sum_{a_i,a_j}
X^\gr{n_i}{a_i}_\gr{m_i}{a_i} \ot X^\gr{n_j}{a_j}_\gr{m_j}{a_j}\right.
\\
\left.
+ \sum_{n_i,n_j,m_i,m_j}
B_{ij}\mbox{}_{n_in_j}^{m_im_j}
\sum_{a_i,a_j,a_i^\prime,a_j^\prime}
R_{ij}\mbox{}^{a_i a_j}_{a_i^\prime a_j^\prime }(x_i/x_j)
 X^\gr{n_j}{a_j^\prime}_\gr{m_i}{a_i}
\ot X^\gr{n_i}{a_i^\prime }_\gr{m_j}{a_j}
\right) ,
\nn
\ena
where the $R$-matrix
\[
R_{\V_{\la_i}\ot \V_{\la_j} }(x_i/x_j) |a_i\rangle\ot|a_j\rangle =
\sum_{a_i^\prime,a_j^\prime}
R_{ij}\mbox{}^{a_i a_j}_{a_i^\prime a_j^\prime }(x_i/x_j)
|a_i^\prime\rangle\ot|a_j^\prime\rangle .
\]
is the intertwining operator between two actions of affine quantum group
$\hg$ on  $\V_{\la_i}\ot \V_{\la_j}$, which are induced
correspondingly by co-multiplication
$\D$ and opposite co-multiplication $\bar{\D}$
\cite{Dr86,J85,J86_1}:
\[
R_{\V_{\la_i}\ot \V_{\la_j} }(x_i/x_j) \D(g)=
\bar{\D}(g) R_{\V_{\la_i}\ot \V_{\la_j} }(x_i/x_j).
\]
$A_{ij}$ and $B_{ij}$, $B_{ii}=0$ in \re{HactMat} are arbitrary matrices.
In general, $H(A,B)$ depends on deformation parameter $q$ of quantum
group, which is included in the $R$-matrix.
Note that  $R_{\V_{\la_i}\ot \V_{\la_j} }(x_i/x_j)$ does not depend
on $q$ and is identity
only if
$\la_i$ or $\la_j$ are trivial one-dimensional representations.
So, in the special
case, when the only nontrivial $R$-matrixes in \re{HactMat} are
between representations, one of which is trivial representation,
the expression of $H(A,B)$ doesn't depend on $q$. Then $H(A,B)$
commutes with the quantum group action for all values of
deformation parameter. In the following  we consider only this case.

Following \cite{RA,HS} we can from the operator $H$ construct
the following Hamiltonian acting on $\W=\V^{\ot L}$:
\footnote{Here and in the following we omit the dependence on $x_i$}
\be
\H=\sum_{i=1}^{L-1}H_{ii+1},
\label{genhamil}
\ee
where the indices $i$ and $i+1$ denote the sites where $H$
acts non-trivially. By the construction, $\H$ is quantum group invariant:
\bea
{[}\H,\D^{L-1}(g){]}=0, & \forall g\in \hg
\nn
\ena
Let us define the projection operators $\Pro^i$ on $\V$ for each class of
equivalent irreps $(\la_i,x_i)$, $i=1,\dots,M$
\bea
\Pro^iv_j=\delta_{ij}v_j, & \forall v_j\in V_{\la_j}(x_j) \nn\\
\sum_{i=1}^M \Pro^i=\id, & (\Pro^i)^2=\Pro^i\nn
\ena
Their action on $\W$ is given by
\[
\Pro^i=\sum_{l=1}^L \Pro^i_l
\]
It is easy to see that these projections commute with Hamiltonian ${\cal H}$
and quantum group $\hg$:
\bea
\label{com}
[\Pro^i,{\cal H}]=0, & [\Pro^i,\hg]=0
\ena
Denote by
$\W_{p_1\dots p_M}$
the subspace of $\W$ with values $p_i$
of $\Pro^i$ on it. Then we have the decomposition
\be
\label{wdecompos}
\W=\bigoplus_{\stackrel{p_1,\dots,p_M}{p_1+\dots+p_M=L}}\W_{p_1\dots p_M}
\ee

Let $\V^0$ be the linear space, spanned by the highest
weight vectors in $V$:
$$
\V^0:=\op_{i=1}^N v_{\la_i}^0,
$$
where  $v^0_{\la_i}\in
\V_{\la_i}$ is a highest weight vector. We also define  $\W^0:=\V^{0\ \ot L}$.
The space
$\W^0$ is $\H$-invariant.
For general $q$ the action of $\hg$ on $\W^0$ generate whole
space $\W$.
Indeed, the $\hg$-action on each state of type
$v^0_{\la_{i_1}}\ot\ldots\ot v^0_{\la_{i_L}}$ generates the space
$\V_{\la_{i_1}}\ot\ldots\ot \V_{\la_{i_L}}$, because the tensor product of
finite dimensional irreducible
representations of an affine quantum group is irreducible \cite{ChP}.

Consider now the subspace $\W^0_{p_1\dots p_M}=\W^0 \cap \W_{p_1\dots p_M}$.
According to (\ref{wdecompos}) we have the decomposition
\be
\label{w0decompos}
\W^0=\bigoplus_{\stackrel{p_1,\dots,p_M}{p_1+\dots+p_M=L}}\W^0_{p_1\dots p_M}.
\ee
Note that
$$
d_{p_1\dots p_M}:=\dim \W^0_{p_1\dots p_M}  =
\left(\begin{array}{c} L \\ p_1\ldots p_M
\end{array}\right)N_1^{p_1}\ldots N_M^{p_M} .
$$

Let us define by ${\cal H}_0$ the restriction of ${\cal H}$ on
$\W_0$: ${\cal H}_0:={\cal H}|_{\W_0}$.
It follows from
(\ref{com})  that  Hamiltonians ${\cal H}$ and ${\cal H}_0$ have
block diagonal form with respect to the decompositions (\ref{wdecompos}) and
(\ref{w0decompos}), respectively.
Every eigenvector
$w^0_{\alpha_{p_1\dots p_M}}\in
\W^0_{p_1\dots p_M}$ with energy value $E_{\alpha_{p_1\dots p_M}}$
gives rise to an irreducible $\hg$-multiplet
$\W_{\alpha_{p_1\dots p_M}}$
of dimension
\be
\label{lev}
\dim \W_{\alpha_{p_1\dots p_M}}=
                        \prod_{k=1}^M(\dim \V_{\la_k})^{p_k}
\ee
On $\W_{\alpha_{p_1\dots p_M}}$
 the Hamiltonian $\H$ is diagonal with eigenvalue
$E_{\alpha_{p_1\dots p_M}}$.
In particular, in the
case when all $\V_{\la_i}$ are equivalent, the degeneracy levels
are  the same for all $E_{\alpha_{p_1\dots p_M}}$ and are equal to
$(\dim \V_\la)^L$.

Now, let us assume we know  the energy spectrum
$E_{\alpha_{p_1\dots p_M}}$ for ${\cal H}_0$.
Then the statistical sum is given by
\be
\label{ZH0}
Z_{{\cal H}_0}(\beta)=\sum_{\stackrel{p_1,\dots,p_M}{p_1+\dots+p_M=L}}
\sum_{\alpha_{p_1\dots p_M}=1}^{d_{p_1\dots p_M}}
\exp(-\beta E_{\alpha_{p_1\dots p_M}}),
\ee
and it follows that the statistical sum of ${\cal H}$ has the following form:
\be
\label{ZH}
Z_{{\cal H}}(\beta)=\sum_{\stackrel{p_1,\dots,p_M}{p_1+\dots+p_M=L}}
\prod_{k=1}^M(\dim \V_{\la_k})^{p_k}
\sum_{\alpha_{p_1\dots p_M}=1}^{d_{p_1\dots p_M}}
\exp(-\beta E_{\alpha_{p_1\dots p_M}}).
\ee
So, if the underlying Hamiltonian ${\cal H}_0$ is integrable and
its eigenvectors and eigenvalues can be found, then
we know these  for ${\cal H}$ too.
Acting with the quantum group on all
eigenvectors of an energy level of $\H_0$ one obtains the whole
eigenspace of ${\cal H}$ for this level.


\section{Multi-band $t-J$ model with vanishing spin-spin coupling $J=0$}

Let us consider here the quantum group $U_q\widehat{sl}_2$.
We choose $\V=\V_0\oplus \V_j$
for decomposition \re{Vdecompos}, i.e.\ we take a direct sum of the trivial
spin-$0$ and the $2j+1$-dimensional spin-$j$ representation of
$U_qsl_2$. The
$R$-matrix in the second term in \re{HactMat} does not depend on $q$ and
spectral parameters $x_i$ and coincides with the identity, as it was
mentioned above. So, using \re{HactMat} and \re{genhamil}, we
obtain the following Hamiltonian
\bea
\label{RA0}
{\cal H}(t,V_1,V_2)=\sum_{i=1}^{L-1}
\left[
-t\sum_{p=1}^{2j+1}
(X_i\mbox{}^p_0 X_{i+1}\mbox{}^0_p+
X_{i+1}\mbox{}^p_0 X_{i}\mbox{}^0_p)
+ V_1 X_{i}\mbox{}^0_0 X_{i+1}\mbox{}^0_0 \right.\nn
\\
\left.+V_2\sum_{p,p^\prime=1}^{2j+1} X_i\mbox{}^p_p
X_{i+1}\mbox{}^{p^\prime}_{p^\prime}
\right]
\ena
The Hamiltonian  ${\cal H}=\sum_i H_{i i+1}$ was constructed
from the  operator $H=H_{i i+1}$, where $H$
can be written in the matrix form
\be
\label{R}
H= \left(
        \begin{array}{cccc}
                 V_1 & 0 & 0 & 0 \\
        0 & 0 & -t\cdot \id & 0\\
                0 & -t\cdot \id & 0 & 0\\
         0 & 0 & 0 & V_2\cdot\id
     \end{array}
\right).
\ee
The projection on the highest weight space coincides with the
constructing block of the $XXZ$ Hamiltonian in an external magnetic
field. This implies that the restriction of \re{RA0} to  the space $\W^0$ is
\bea
\label{XXZ}
\H_0(t,W_1,W_2)&=&\H_{XXZ}(t,\Delta,B)
\nn\\
&=&-\frac{t}{2}\sum_{i=1}^{L-1} \left(
\sigma^x_i \sigma^x_{i+1}+\sigma^y_i \sigma^y_{i+1} +
\Delta\sigma^z_i \sigma^z_{i+1} +\frac B2\sigma^z_i\right),
\ena
where
\be
\label{DeltaW}
\Delta=-\frac {V_1+V_2}{2t}, \qquad
B=\frac 2t(V_1-V_2)
\ee
For the special case $V_1+V_2=0$ $\H_0$ gives rise to the free
fermionic (or equivalently $XY$) Hamiltonian ($\Delta=0$).

The projection operators $X^a_b$ are expressed through the
fermionic creation-an\-ni\-hi\-la\-tion operators as follows
\bea
\label{proj2}
 X_i\mbox{}^p_0= \Proj c_{i,p}^+, \qquad &
 X_i\mbox{}_p^0=c_{i,p}\Proj, \nn\\
X_i\mbox{}^p_p=n_{i,p}\Proj=\Proj n_{i,p}, &
X_i\mbox{}^0_0= (1-n_{i})\Proj=\Proj (1-n_i)
\ena
Here we introduced the projection operator which forbids double occupation
on all sites
$$
 \Proj=\prod_{i=1}^L { \Proj}_i, \qquad 
\Proj_i=\prod_{p\ne p^\prime}
(1-n_{i,p}n_{i,p^\prime})
$$
and the total particle number $n_i=\sum_pn_{i,p}$ at site $i$.

After the substitution of the fermionic representation \re{proj2}
into \re{RA0} we obtain
\bea
\label{RA1}
{\cal H}(t,V_1,V_2) = {\cal P}\sum^{L-1}_{i=1} \left[ -t
\sum_{p=1}^{2j+1}
 ( c^{+}_{i,p}
c_{i+1,p} + c^{+}_{i+1,p}c_{i,p}  )\right.
+\nn\\
\left. V n_in_{i+1}-V_1(n_i+n_{i+1})+V_1\right] {\cal P},
\ena
where $V=V_1+V_2$.
The chemical potential term $-V_1\sum_{i=1}^{L-1}(n_i+n_{i+1})$
commutes with ${\cal H}$ and can be omitted. So, up to
unessential boundary and constant terms \re{RA1} is a multicomponent
$t-J$ model with vanishing spin-spin coupling ($J=0$)
\bea
\label{RA}
{\cal H}(t,V) = \sum^{L-1}_{i=1} \left[ -t
\sum_{p=1}^{2j+1}
 ( c^{+}_{i,p}c_{i+1,p} + c^{+}_{i+1,p}c_{i,p}  ) + Vn_in_{i+1} \right]
+\sum_{i=1}^L \sum_{\stackrel{p\ne p^\prime,}{p,p^\prime=1}}^{2j+1} 
U_{p,p^\prime}
n_{i,p}n_{i,p^\prime},
\ena
where the infinite Hubbard interaction amplitude
$U_{p,p^\prime}=+\infty$
 between  $p$ and $p^\prime$ bands excludes sites with double and
 more occupations.
 It follows from the above considerations that  this model has
 energy levels which  coincide with the levels of $XXZ$ Heisenberg
 model, but that the degeneracy of the levels is different.

For vanishing density-density interaction $V=0$ the Hamiltonian
\re{RA} describes the infinite repulsion limit of the multi-band
Hubbard model. Thus, according to \re{DeltaW} $\Delta=0$
and it has the energy levels of the free fermionic model.

\section{Multi-band extension of the  $t-J$ model with affine
quantum group symmetry}

In this section we consider Hamiltonians which have the
same energy levels as $t-J$ model but have affine quantum group
symmetry. Because each site in ordinary
$t-J$ model has three states, one should for this purpose take
direct sum of three spaces. Let
\be
\label{t-J-de}
\V=\V_0\oplus \V_{j}\oplus \V_{j}
\ee
Recall that the $t-J$ model is given by
\be
\label{t-J}
\H_{t-J}(t,J,V)=
{\cal P}\sum^{L-1}_{i=1} \left[ -t
\sum_{\sigma=\pm\frac12}
 ( c^{+}_{i,\sigma}
c_{i+1,\sigma} + c^{+}_{i+1,\sigma}c_{i,\sigma}  ) +J \Sp_i\Sp_{i+1}
+V n_in_{i+1}\right] {\cal P},
\ee
where $c^+_{\sigma}, \ c_\sigma$ are creation-annihilation operators of 
spin-$\frac12$ fermion, $\Sp=\sum_{\sigma,\sigma^\prime}
c^+_\sigma {\bf\sigma}_{\sigma\sigma^\prime}c_{\sigma^\prime}$
is the fermionic spin operator and ${\cal
P}=\prod_{i=1}^L(1-n_{i,\up}n_{i,\dow})$ forbids  double
occupation of sites.

We  rewrite it in terms of Hubbard operators $X^a_b$, where $a,b=0,\pm\frac12$:
\bea
\label{t-J-X}
{\cal H}(t,J,V)=\sum_{i=1}^{L-1}
\left[
\sum_{\sigma=\pm\frac12}\left(-t
(X_i\mbox{}^\sigma_0 X_{i+1}\mbox{}^0_\sigma+
X_{i+1}\mbox{}^\sigma_0 X_{i}\mbox{}^0_\sigma)
+ \frac 12 J\cdot X_{i}\mbox{}^\sigma_{-\sigma}
X_{i+1}\mbox{}^{-\sigma}_{\sigma} \right)\right.\nn
\\
\left.
+\sum_{\sigma,\sigma^\prime=\pm\frac12}
\left({\sigma\sigma^\prime} J+V\right)
X_i\mbox{}^\sigma_\sigma
X_{i+1}\mbox{}^{\sigma^\prime}_{\si^\pr}
\right]
\ena
Let us now look at the  general expression \re{HactMat} of intertwining
operators $H_{ij}$ acting on the space \re{t-J-de}.
For convenience we make index change in the following way.
The two spin-$j$ representations we use are denoted by $\sigma=\pm\frac12$.
The intrinsic index in each $V_j^{(\si)}$
is denoted  by $k$, $k=1,\dots,2j+1$.
So, instead of $(n_i,a_i)$ in \re{HactMat}
we have $(\si,k)$, if $i$ corresponds to  spin-$j$ multiplet.
Because the
spin-$0$ singlet is one dimensional and single, we just use for it the
index $0$.  The non-equivalent irreps in \re{t-J-de} are $V_j^{(\si)}$
and $V_0$ and, as mentioned above, the $R$-matrix for two such representations
 is the identity.  After performing the first sum in
\re{HactMat} over non-equivalent multiplets
we obtain
\bea
\label{HactMat-t-J}
H(A,a,b_1,b_2)=
\sum_{\si_1,\si_2,\si_1^\pr,\si_2^\pr}
\left(
A_{}\mbox{}_{\si_1\si_1^\pr}^{\si_2\si_2^\pr}
\sum_{k,k^\pr}
X^\gr{\si_1}{k}_\gr{\si_2}{k} \ot
X^\gr{\si_1^\pr}{k^\pr}_\gr{\si_2^\pr}{k^\pr}
\right)+ a\cdot X^0_0\ot X^0_0
\nn   \\
+\sum_{k,\sigma}
 \left(b_1 \cdot
X^\gr{\si}{k}_{0} \ot X^{0 }_\gr{\si}{k}
+b_2 \cdot X^0_\gr{\si}{k} \ot X^\gr{\si}{k}_0
\right)
\ena
To implement the  restriction $H_0(A,a,b_1,b_2)$ of this operator
on the highest weight space one just should eliminate the sum over $k,k^\prime$
 and put $k=k^\prime=0$. Comparing \re{HactMat-t-J} and \re{t-J-X} it follows 
that
the expressions coincide if one chooses
\[
a=0 \qquad b_1=b_2=-t \qquad
A_{\si-\si}^{-\si\si}=J/2
\qquad
A_{\si\si}^{\si^\pr\si^\pr}=(\si\si^\pr)\cdot J+V
\]
and choose the other values of $A_{\si_1\si_2}^{\si_1^\pr\si_2^\pr}$
equal zero.

So, the Hamiltonian $\H(A,a,b_1,b_2)$ corresponding to
\re{HactMat-t-J} with these values of parameters
gives rise to a $t-J$ model \re{t-J}  on the highest weight space.
According to the previous considerations it will have the same energy
levels as $t-J$ model, but  with different degeneracy. Recall that
for $J=2t$ the $t-J$ model is "supersymmetric" and integrable.

We express the Hubbard operators in terms of multi-band fermionic
creation\--an\-ni\-hi\-la\-tion operators
as follows
\bea
\label{proj3}
 X_i\mbox{}^\gr{\sigma}{k}_0=\Proj c_{i,\sigma}^{k+}, \qquad &
 X_i\mbox{}_\gr{\sigma}{k}^0=c_{i,\sigma}^k\Proj ,  \nn\\
\\
 X_i\mbox{}^\gr{\sigma}{k}_\gr{-\sigma}{k}
=c_{i,\sigma}^{k+}c_{i,-\sigma}^k\Proj=
\Proj c_{i,\sigma}^{k+}c_{i,-\sigma}^k,
& X_i\mbox{}^\gr{\sigma}{k}_\gr{\sigma}{k}
=n_{i,\sigma}^k\Proj=\Proj n_{i,\sigma}^k
\nn 
\ena
Here as before we used the projection operator, which forbids double occupation
on all sites
$$
 \Proj=\prod_{i=1}^L { \Proj}_i, \qquad 
\Proj_i=\prod_{\gr{\sigma}{k}\ne \gr{\sigma^\prime}{k^\prime}}
(1-n_{i,\sigma}^k n_{i,\sigma^\prime}^{k^\prime})
$$
Now, we can write down the
Hamiltonian \re{genhamil} in terms of
multi-band fermions, substituting \re{proj3} into \re{HactMat-t-J}.
We obtain in this way the multi-band
generalization of \re{t-J}
\be
\label{Mu-t-J}
\H(t,J,V)=
{\cal P}\sum^{L-1}_{i=1} \left[ -t
\sum_{k=1}^{2j+1}\sum_{\sigma=\pm\frac12}
 ( c^{k+}_{i,\sigma}
c^k_{i+1,\sigma} + c^{k+}_{i+1,\sigma}c^k_{i,\sigma}  ) +J
\Sp_i\Sp_{i+1}
+V n_in_{i+1}\right] {\cal P},
\ee
Here $k$ is the band index, and $\Sp=\sum_k\Sp^k$, $n=\sum_k
n^k$ are total spin and total particle number operators.It's
easy to see that we have conservation of the particle number
operators $\sum_{i,\sigma}^{k} = n^k$ for the all colours $k$. 
\section{Bethe ansatz for the $t-J$ model with affine quantum group symmetry}

The goal of this section is to apply the  Bethe Ansatz technique to
our model and derive the corresponding Bethe equations for the excitations.

After making some trivial Pauli matrix calculations one can represent
Hamiltonian \re{Mu-t-J} as
\bea
\label{25}
\H(t,J,V)=
{\cal P}\sum^{L-1}_{i=1} \left[ -t
\sum_{k=1}^{2j+1}\sum_{\sigma=\pm\frac12}
 ( c^{k+}_{i,\sigma}
c^k_{i+1,\sigma} + c^{k+}_{i+1,\sigma}c^k_{i,\sigma}  ) + \right.\nn\\
\left.2J
\sum_{k,k^\prime,\sigma\ne \tau} c^{k+}_{i,\sigma}c^k_{i,\tau}
c^{k^\prime+}_{i+1,\tau}c^{k^\prime}_{i+1,\sigma} + 
(V-J)n_in_{i+1}\right] {\cal P}.
\ena
Due to the conservation of the particle number operator $n^k$, and
according to the coordinate Bethe Ansatz we look for eigenvectors
of \re{25}, corresponding to $N$ fermions of $2j+1$ bands in the
following form
\be
\label{26} 
|\Psi> =\sum_{k_1...k_N =1}^{2j+1} \sum_{x_{1}\sigma_{1}}...\sum_{x_N\sigma_{N}}
\psi^{k_1...k_N}(x_1\sigma_1,...,x_N\sigma_N)c^{k_1+}_{x_1\sigma_1}...
c^{k_N+}_{x_N\sigma_N}|0>,
\ee
where $|0>$ is empty vacuum state.

The eigenvalue equation $H |\Psi> = E|\Psi>$ in the one particle sector
\be
\label{27}
-t(\Psi^k(x-1,\sigma)+\Psi^k(x,\sigma)) = E\Psi^k(x,\sigma)
\ee
gives us
\be
\label{28}
E=-2 t \cos p
\ee
after substituting of the plane wave function with momentum p into \re{27}.

The eigenvalue equations in the two particle sector  allows us to fix
the energy of the state as a sum of two one particle energies, as well
as the two two particle scattering matrix $S^{\sigma^\prime_1\sigma^\prime_2,
k^\prime_1k^\prime_2}_{\sigma_1\sigma_2, k_1 k_2}(p_1,p_2)$. We choose the
antisymmetric wave function $\Psi^{k_1 k_2}(x_1 \sigma_1,x_2 \sigma_2)$
as 
\be
\label{29}
\psi^{k_1, k_2}(x_1\sigma_1, x_2\sigma_2)=
A^{k_1 k_2}(p_1\sigma_1,p_2\sigma_2)e^{i(p_1x_1+p_2x_2)}
- A^{k_2 k_1}(p_2\sigma_1,p_1\sigma_2)e^{i(p_2x_1+p_1x_2)}
\ee
for $x_2 \leq x_1$ and
\be
\label{30}
\psi^{k_1, k_2}(x_1\sigma_1, x_2\sigma_2)=
A^{k_2 k_1}(p_2\sigma_2,p_1\sigma_1)e^{i(p_1x_1+p_2x_2)}
- A^{k_1 k_2}(p_1\sigma_2,p_2\sigma_1)e^{i(p_2x_1+p_1x_2)}
\ee
for $x_1 \leq x_2$.

The continuity condition at $x_1=x_2$  should be imposed:
\be
\label{31}
A^{k_1 k_2}(p_1\sigma_1,p_2\sigma_2)- A^{k_1 k_2}(p_2\sigma_1,p_1\sigma_2)
=A^{k_2 k_1}(p_2\sigma_2,p_1\sigma_1)- A^{k_2 k_1}(p_1\sigma_2,p_2\sigma_1)
\ee
 Use of the Hamiltonian \re{Mu-t-J} and the most general form
\re{26} of the eigenfunctions $\psi$, the eigenvalue equations can be written as
\bea
\label{32}
 -t\left[\psi^{k_1 k_2}(x_1+1 \sigma_1, x_2 \sigma_2) (1-\delta_{x_1+1,x_2})+
\psi^{k_1 k_2}(x_1-1 \sigma_1, x_2 \sigma_2) (1-\delta_{x_1,x_2+1})+ 
\right.\nn\\
\left.\psi^{k_1 k_2}(x_1 \sigma_1, x_2+1 \sigma_2) (1-\delta_{x_1,x_2-1})+t
+ \psi^{k_1 k_2}(x_1 \sigma_1, x_2-1 \sigma_2) 
(1-\delta_{x_1-1,x_2})\right]+\nn\\
2J \delta_{|x_1-x_2|, 1}\psi^{k_1 k_2}(x_1 \sigma_2, x_2 \sigma_1)+ 
+ (V-J) \delta_{|x_1-x_2|, 1} \psi^{k_1 k_2}(x_1 \sigma_1, x_2 \sigma_2) \nn\\
= E \psi^{k_1 k_2}(x_1 \sigma_1, x_2 \sigma_2)
\ena

The terms $(1-\delta)$ appeared near the hopping terms as a result of projective
operator $\cal P$, preventing double occupancy of the sites.

Two different cases can be considered:
\begin{itemize}
\item[I.]
 $\;\; |x_1-x_2|>1$. In this case eigenvalue equation \re{32} is reduced to
\bea
\label{33}
-t[\psi^{k_1 k_2}(x_1+1 \sigma_1, x_2 \sigma_2)+
\psi^{k_1 k_2}(x_1-1 \sigma_1, x_2 \sigma_2)+
\psi^{k_1 k_2}(x_1 \sigma_1, x_2+1 \sigma_2)+\nn\\
 + \psi^{k_1 k_2}(x_1 \sigma_1, x_2-1 \sigma_2)]
=E \psi^{k_1 k_2}(x_1 \sigma_1, x_2 \sigma_2).
\ena

After substitution of the expression \re{29}-\re{30} for the plane
waves into \re{33} and some simple calculations, the spectrum of
two particle state can be found.
\be
\label{34}
E=-2t (\cos p_1 + \cos p_2).
\ee

\item[II.] $\;\; |x_1-x_2| =1$. Without loos of generality one can take 
$x_2=x_1+1$.
Then the eigenvalue equation reduces to
\bea
\label{35}
-t[\psi^{k_1 k_2}(x_1-1 \sigma_1, x_2 \sigma_2)+
\psi^{k_1 k_2}(x_1 \sigma_1, x_2+1 \sigma_2)+
2J \psi^{k_1 k_2}(x_1 \sigma_2, x_2 \sigma_1)+\nn\\
+ (V-J) \psi^{k_1 k_2}(x_1 \sigma_1, x_2 \sigma_2)=
E \psi^{k_1 k_2}(x_1 \sigma_1, x_2 \sigma_2)
\ena
By substitution of \re{29}  and use of  continuity conditions \re{31}
one can express the amplitude
$A^{k_2 k_1}(p_2 \sigma_1, p_1 \sigma_2$) after scattering  
through $A^{k_1 k_2}(p_1 \sigma_1, p_2 \sigma_2)$ and 
$A^{k_1 k_2}(p_1 \sigma_2, p_2 \sigma_1)$ before, and, therefore get R-matrix
of the model:
\be
\label{36}
A^{k_2 k_1}(p_2 \sigma_1, p_1 \sigma_2) = R^{\sigma^{'}_{1} 
\sigma^{'}_{2},k_2 k_1}_{\sigma_{1} \sigma_{2}, k^{'}_1 k^{'}_2}
A^{k^{'}_1 k^{'}_2}(p_1 \sigma_1^{'}, p_2 \sigma_{2}^{'})
\ee
where $R$-matrix is the product of ordinary $t-J$ model $R$-matrix
multiplied the permutation operator in the $k$-index space.
\be
\label{37}
R^{\sigma^{'}_{1} 
\sigma^{'}_{2},k_2 k_1}_{\sigma_{1} \sigma_{2}, k^{'}_1 k^{'}_2} =
R^{\sigma^{'}_{1} \sigma^{'}_{2}}_{\sigma_{1} \sigma_{2}}(t-J) \cdot 
\delta^{k_2}_{k^{'}_1} \cdot \delta^{k_1}_{k^{'}_2}.
\ee
At the  supersymmetric point $2J=t,\;V=-J/4$
\be
\label{38}
R^{\sigma^{'}_{1} \sigma^{'}_{2}}_{\sigma_{1} \sigma_{2}}(t-J)=
\frac {(\lambda_{1} - \lambda_{2}) \hat{P} + i \hat{I}}{\lambda_{1}-\lambda_{2} 
+ i}.
\ee
In \re{38} $\lambda =\frac{1}{2} \cot\frac{p}{2}$ is the rapidity, 
$\hat{I}= \delta^{\sigma^{'}_{1}}_{\sigma_{1}}\delta^{\sigma^{'}_{2}}_
{\sigma_{2}}$ ,and 
$\hat{P}= \delta^{\sigma^{'}_{2}}_{\sigma_{1}}\delta^{\sigma^{'}_{1}}_
{\sigma_{2}}$.
\end{itemize}

The scattering matrix $S^{\sigma^{'}_{1} 
\sigma^{'}_{2},k^{'}_1 k^{'}_2}_{\sigma_{1} \sigma_{2}, k_1 k_2}$
will be  defined multiplying the $R$-matrix by permutation operator 
$\hat{P}$ in the spin($\sigma$) and colour($k$) spaces.

The exact integrability of the model is connected with the fact that the 
$S$-matrix should fulfill the Yang-Baxter triangular relations
\bea
\label{39}
S^{\sigma^{'}_{1} 
\sigma^{'}_{2},k^{'}_1 k^{'}_2}_{\sigma_{1} \sigma_{2}, k_1 k_2}(\lambda_{1}-
\lambda_{2})\cdot S^{\sigma^{''}_{1} 
\sigma^{'}_{3},k^{''}_1 k^{'}_3}_{\sigma^{'}_{1} \sigma_{3}, k^{'}_1 k_3}
(\lambda_{1}- \lambda_{3})\cdot S^{\sigma^{''}_{2} 
\sigma^{''}_{3},k^{''}_2 k^{''}_3}_{\sigma^{'}_{2} \sigma^{'}_{3}, 
k^{'}_2 k^{'}_3}(\lambda_{2}- \lambda_{3}) = \nn\\
S^{\sigma^{'}_{2} 
\sigma^{'}_{3},k^{'}_2 k^{'}_3}_{\sigma_{2} \sigma_{3}, k_2 k_3}(\lambda_{2}-
\lambda_{3})\cdot S^{\sigma^{'}_{1} 
\sigma^{''}_{3},k^{'}_1 k^{''}_3}_{\sigma_{1} \sigma^{'}_{3}, k_1 k^{'}_3}
(\lambda_{1}- \lambda_{3})\cdot S^{\sigma^{''}_{1} 
\sigma^{''}_{3},k^{''}_1 k^{''}_2}_{\sigma^{'}_{1} \sigma^{'}_{2}, 
k^{'}_1 k^{'}_2}(\lambda_{1}- \lambda_{2})
\ena 
and constraints on $t,J,V$ are imposed just by these equations.

Consider now $N$ itinerant electrons in a box of $L$ sites with periodic
boundary  conditions. If we successively make change  of position  of an
 electron and its neighbouring electron in a chain, each
interchange produces a scattering matrix and  when the particle comes back
to its starting position from the other side, we will have the cyclic
product of $S$-matrices, which is called the Transfer Matrix:
\bea
\label{40}
\hat{T}_{j}(\lambda_{j})=\hat{S}_{j,j+1}(\lambda_{j}-\lambda_{j+1})....
\hat{S}_{j,N}(\lambda_{j}-\lambda_{N})\cdot
\hat{S}_{j,1}(\lambda_{j}-\lambda_{1})
....\hat{S}_{j,j-1}(\lambda_{j}-\lambda_{j-1})
\ena

Here we skip the matrix indexes $\sigma$ and $k$ while the hat on $S$ means the
operator in that space. The periodicity means that the Transfer Matrix  has
to be diagonal for all $j=1.....N$ with the eigenvalues $exp(i p_j N)$, or,
on terms of rapidity $\lambda_{j}$
\be
\label{41}
e^{i p_j N}= \left(\frac{\lambda_{j}+i/2}{\lambda_{j}-i/2}\right)^{L}
\ee

The matrix \re{40} is the trace of so called monodromy matrix, which by
definition
is the product of the $S$-matrices without taking trace. Hence, the
monodromy matrix can be considered as a $(2\times 2)\ot 
((2j+1)\times (2j+1))$ matrix in
the spin $\sigma$ and colour $k$ spaces. Let's   remember  now that this
operator is a  unity operator in the colour space. We will not describe here
the details of the algebraic Bethe Ansatz(see e.g.  \cite{KI}), 
but already now it is clear, that because the monodromy
matrix of our model is the monodromy matrix of the ordinary $t-J$ model
multiplied by the unity matrix in the $k$-space, the generalization of the 
algebraic Bethe Ansatz to the present model give rise to  the same 
equations as the ordinary $t-J$ model.

Specifically we get:
\be
\label{42}
\left(\frac{\lambda_{j}+i/2}{\lambda_{j}-i/2}\right)^{L}=
\prod_{\beta=1}^{M} \frac{\lambda_{j}-\Lambda_{\beta}+i/2}{\lambda_{j}-
\Lambda_{\beta}-i/2},\;\;\;\;\;\;\;\;\;\; j=1,...N
\ee
\be
\label{43}  
\prod_{j=1}^{N} \frac{\Lambda_{\alpha}-\lambda_{j}+i/2}{\Lambda_{\alpha}-
\lambda_{j}-i/2}= - \prod_{\beta=1}^{M}
\frac{\Lambda_{\alpha}-\Lambda_{\beta}+i}
{\Lambda_{\alpha}-\Lambda_{\beta}-i},\;\;\;\;\;\;\;\;\;\alpha=1,...M
\ee
where $L$ is the number of lattice sites, $N$-is the number of electrons
and $M$ is the number of  spin down electrons. 

We see that colour is disappeared
from the equations, which means that we can make arbitrary partition
of colour charges on the state of $N$  particles in a chain and all
wave vectors will become eigenvalues of the Hamiltonian, provided
that their $\lambda$'s fulfill the Bethe equations.

Equations \re{42}-\re{43} are Lai's \cite{L} form of Bethe equations,
written on a basis of empty background.

The total energy is given by
\be
\label{44}
E= -2N + 2 \sum_{j=1}^{N}\frac{1/2}{\lambda^{2}_{j}+1/4}
\ee

However  for the construction of the thermodynamic limit of our model
it is more convenient to use the Sutherland's form \cite{SU} of the Bethe
equations, which  is equivalent to  Lai's equations \cite{EK,BCF}.
In Sutherland's representation one start with a ferromagnetic pseudo-vacuum 
state with all spins up and consider excitations as $N_h$-holes(empty sites)
and $M$ spin down electrons.

In Sutherland's representation we have
\bea
\label{45}
\left(\frac{\lambda_{j}+i/2}{\lambda_{j}-i/2}\right)^{L}=
\prod_{\beta=1}^{M} \frac{\lambda_{j}-\Lambda_{\beta}+i/2}{\lambda_{j}-
\Lambda_{\beta}-i/2},\;\;\;\;\;\;\;\;\;\; j=1,...M+N_h \nn\\
1=\prod_{j=1}^{M+N_h} \frac{\lambda_{j}-\Lambda_{\beta}+i/2}{\lambda_{j}-
\Lambda_{\beta}-i/2},\;\;\;\;\;\;\;\;\;\;\;\;\;\;\;\;\;\beta=1,...M
\ena
The equations \re{45} have real and complex solutions. The complex solutions
are in a form known as strings, which may be found by fixing $N_h$ and $M$
(they are conserved quantities) and letting the lattice size 
$L\rightarrow \infty$.
Following \cite{EK,W} one finds complex solutions in the form
\bea
\label{46}
\lambda_{j}=\lambda + i (n+1-2j),\;\;\;\;\;\;\;\;\;\; j=1,...n\nn\\
\Lambda_{\tau}=\lambda + i (n-2 \tau)\;\;\;\;\;\;\;\;\;\;\tau=1,...n-1,
\ena 
for arbitrary $n$. In a finite box these string solutions are not exact but
as in \cite{L,T,G} we will assume that the corrections 
are small in power of $L$ and hence vanish in the thermodynamic limit.

We would like to write the equation for the centrum of the strings of length
$n$, $\lambda^{n}_{\alpha}$, (which are the real variables), let their
number is $M^{'}$ and rapidities
$\Lambda_{\beta}$ of spin down electrons.

Substituting \re{45} into Lai's form of Bethe equations and taking the 
logarithms from the left and right hand sides, one can get
\bea
\label{47}
L \theta(\frac{\lambda^{n}_{\alpha}}{n})=2\pi I^n_{\alpha}+\sum_{m=1}^{\infty}
\sum_{\gamma}^{N_m}\Theta_{nm}(\lambda_{\alpha}^{n}-\lambda_{\gamma}^{m})
+\sum_{\beta=1}^{M}\theta(\frac{\lambda_{\alpha}^{n}-\Lambda_{\beta}}{n})
\ena
and
\bea
\label{48}
\sum_{n=1}^{\infty}\sum_{\alpha=1}^{N_n}\theta(\frac{\Lambda_{\beta}-\lambda_{\a
lpha}
^{n}}{n}) = 2 \pi J_{\beta}
\ena
where $N_m$ is the number of strings of length $m$, $\theta(x)=2tan^{-1}(x)$
and
\be
\label{49}
\Theta_{mn}=\left\{\begin{array}{l}
\theta(\frac{x}{|n-m|})+2\theta(\frac{x}{|n-m|+2})
+...+2\theta(\frac{x}{n+m-2})\;\;\;\;\;\;if\;\;n\ne m \nn\\
2\theta(\frac{x}{2})+2\theta(\frac{x}{4})+...+
2\theta(\frac{x}{n+m-2})\;\;\;\;\;\;\;\;\;\;\;\;\;\;\;\;\;\;\;if\;\;n=m
\end{array}\right.
\ee

$I_{\alpha}^{n}$ and $J_{\beta}$ are integers(half integers)
appearing after choice of branch of the logarithm.

The integers $M$, $N_m$ and $M^{'}$ satisfy the relation
\be
\label{50}
M^{'}=M+\sum_{m=1}^{\infty}mN_{m}
\ee
Any solution is defined by a set of $I_{\alpha}^{n}$ and $J_{\beta}$. All
possible integer(half integer) values defines the states called vacancies.
The vacancies may be occupied by particles of colour $k$ or not occupied
at all. In a case of occupied vacancies one can mark the corresponding 
$I_{\alpha}^{n}$ and $J_{\beta}$ by colour index $k$ as for particles,
$I_{\alpha}^{n,k}$ and $J_{\beta}^{k}$. When the state is empty, the 
corresponding integer will be mentioned as $\bar I_{\alpha}^{n}$ and 
$\bar J_{\beta}$.

One may calculate the number of string states \re{46} taking account also
colour degrees of freedom and will get $(2j+1)^{L}$.

\section{The thermodynamic limit of the Bethe Ansatz equations}
 
In the thermodynamic limit $L\rightarrow \infty$  the solution becomes densely
packed (the difference of two neighbour solutions is of order $1/L$, 
$\lambda_{j+1}^{n}-\lambda_{j}^{n}=O(1/L)$) and one can introduce
density functions of the states and pass from sum's to integrals in 
the equations \re{47}-\re{48}.

Let's define the particle and hole densities as follows:
\bea
\label{51}
\rho_{p}^{n,k}(\lambda)=\lim_{L\rightarrow\infty}\frac{1}
{L(\lambda_{I_{j+1}^{n,k}}-\lambda_{I_{j}^{n,k}})},
\;\;\;\;\;\;\;\rho_{h}^{n}(\lambda)=\lim_{L\rightarrow\infty}\frac{1}
{L(\lambda_
{\bar I_{j+1}^{n}}-\lambda_{\bar I_{j}^{n}})}\nn\\
\sigma_{p}^{k}(\lambda)=\lim_{L\rightarrow \infty}\frac{1}{L(
\Lambda_{J_{j+1}^{k}}-\Lambda_{J_{j}^{k}})},
\;\;\;\;\;\;\;\sigma_{h}(\lambda)=\lim_{L\rightarrow \infty}\frac{1}
{L(\Lambda_{\bar J_{j+1}}-\Lambda_{\bar J_{j}})}
\ena
and the sum of the hole and particle densities defines the density
functions of vacancies.
\be
\label{52}
\rho_{t}^{n}=\sum_{k=1}^{2j+1}\rho_{p}^{n,k}+\rho_{h}^{n},\;\;\;\;\;\;\;\;\;\;
\sigma_{t}=\sum_{k=1}^{2j+1}\sigma_{p}^{k}+\sigma_{h}.
\ee

The passage from  sum to integral is straightforward and in the thermodynamic
limit the Bethe equations \re{47}-\re{48} become integral equations:
\be
\label{53}
\rho_{t}^{n}(\lambda)=f_{n}(\lambda)-\sum_{m=1}^{\infty}\int_{-\infty}^{\infty}
d\lambda^{'}A_{n,m}(\lambda-\lambda^{'})\rho_{p}^{m}(\lambda^{'})-
\int_{-\infty}^{\infty}
d\Lambda f_{n}(\lambda-\Lambda)\sigma_{p}(\Lambda)
\ee
and
\be
\label{54}
\sigma_{t}(\Lambda)=\sum_{n=1}^{\infty}\int_{-\infty}^{\infty} d\lambda 
f_{n}(\Lambda-\lambda)\rho_{p}^{n}(\lambda)
\ee
 $f_{n}(\lambda)$ and $A_{n,m}(\lambda)$ in equations \re{53}-\re{54} are
defined by
\be
\label{55}
f_{n}(\lambda)=\frac{1}{2 \pi}\frac{d \theta(\lambda)}{d \lambda}=
\frac{1}{\pi}\frac{n}{\lambda^{2}+n^2}
\ee
and
\be
\label{56}
A_{n,m}=\left\{\begin{array}{l}f_{|n-m|}(\lambda)+2f_{|n-m|+2}(\lambda)+
...+2 f_{n+m-2}(\lambda),\;\;\;\;\;\;\;\;\;\;if \;\;\;n \ne m\nn\\
2f_{2}(\lambda)+2f_{4}(\lambda)+...+2f_{2n-2}(\lambda),\;\;\;\;\;\;\;\;
\;\;\;\;\;\;\;\;\;\;\;\;\;\;\;\;\;\;\;\;\;if\;\;\;n=m
\end{array}\right.
\ee

Following Yang and Yang \cite{YY}, Takahashi \cite{T} and Gaudin
\cite{G}, we can write down the thermodynamic equilibrium equations. 
For the ordinary $t-J$ 
model it was done in \cite{S1,JK,W}.

The conserved quantities of our model are:
\begin{itemize}
\item
{\it the energy}
(the expression in ferromagnetic background slightly differs from
the Lai's expression \re{44})
\be
\label{57}
E=L\left(1-4 \pi \sum_{n=1}^{\infty}\sum_{k=1}^{2j+1}\int d\lambda 
f_{n}(\lambda)\rho_{p}^{n,k}(\lambda)\right),
\ee 
\item
{\it the number of different particles}
\be
\label{58}
N^k = L\left(1-\int d\lambda \sum_{n=1}^{\infty}\rho_{p}^{n,k}(\lambda)+
\int d\Lambda \sigma_{p}^{k}(\Lambda)\right)
\ee
\item
and {\it the magnetisation}
\be
\label{59}
S_{z}=\frac{L}{2}\left(1-\sum_{n=1}^{\infty}\sum_{k=1}^{2j+1}(2n-1)
\int d\lambda \rho_{p}^{n,k}(\lambda)-\sum_{k=1}^{2j+1}\int d\Lambda
\sigma_{p}^{k}(\Lambda)\right).
\ee
\end{itemize}
Correspondingly, for the free energy $F$ one can write
\be
\label{60}
F=E - \sum_{k=1}^{2j+1}\mu_{k}N^{k} - B S_{z} - T S,
\ee
where $\mu_{k}$ are the chemical potentials of particles, $B$ is the
external magnetic field and $T$ is the temperature.  We
will calculate the entropy $S$ in a standard way, as the logarithm of 
the number of possible 
states of the model in the interval $d\lambda$.
\be
\label{61}
N(\lambda,d \lambda)= e^{S(\lambda)d\lambda}=\frac{L(\sum_{k=1}^{2j+1}
\rho_{p}^{k}d\lambda)!}{\prod_{k=1}^{2j+1}(L\rho_{p}^{k}d \lambda)! 
(L\rho_{h}d\lambda)!}
\ee 
By Stirling's formula one finds from \re{61}
\bea
\label{62}
S=L \left\{ \int d\lambda \sum_{m=1}^{\infty} \left[ \rho_{t}^{n}\
log{\rho_{t}^{n}}-\sum_{k=1}^{2j+1}\rho_{p}^{n,k}\log{\rho_{p}^{n,k}}-
\rho_{h}^{n}\log{\rho_{h}^{n}} \right]+ \right.\nn\\
\left.+\int d\Lambda \left[\sigma_{t}\log{\sigma_{t}}-
\sum_{k=1}^{2j+1}\sigma_{p}^{k}\log{\sigma_{p}^{k}}-
\sigma_{h}\log{\sigma_{h}} \right] \right\}.
\ena
To obtain the equilibrium equations one should minimise the free
energy over density functions of the particles and holes. Puting 
$\delta F =0$ for variations of the  density functions
$\rho_{p}^{n,k}, \rho_{h}^{n}, \sigma_{p}^{k}, \sigma_{h}$ one gets
 after some algebraic transformations an
infinite set of integral equations for the densities. If one defines the 
excited energies as usual
\be
\label{63}
\frac{\rho_{p}^{n,k}}{\rho_{t}^{n}-\rho_{p}^{n,k}}=e^{-\frac{\epsilon^{n,k}}
{T}},\;\;\;\;\;\;\;\;\;\;\;\;\frac{\sigma_{p}^{k}}{\sigma_{t}-\sigma_{p}^{k}}
=e^{-\frac{\epsilon_{\Lambda}^{k}}{T}}
\ee
and
\be
\label{64}
\epsilon_{0}^{n,k}= -4 \pi f_{n}(\lambda)-(2n-1)B + \mu_{k}
\ee
we obtain
\bea
\label{65}
\epsilon^{n,k}(\lambda)=\epsilon_{0}^{n,k}+ T \sum_{m=1}^{\infty}
\int d\lambda^{'} A_{n,m}(\lambda-\lambda^{'}) \log{(1+
e^{\frac{-\epsilon^{m,k}(\lambda^{'})}{T}})}-\nn\\
-T \int d\Lambda f_{n}(\lambda-\Lambda)\log{(1+
e^{\frac{-\epsilon_{\Lambda}^{k}(\Lambda)}{T}})}
\ena
for $n=1,....\infty$, and
\be
\label{66}
\epsilon_{\Lambda}^{k}= -\mu_{k} -B +
T \sum_{n=1}^{\infty} \int d\lambda f_{n}(\lambda-\Lambda)\log{(1+
e^{\frac{-\epsilon^{n,k}(\lambda)}{T}})}.
\ee

It is  seen  from these equations for densities and excitation energies
that if all particles have the same chemical potential $\mu=\mu_{k}$,
the solutions will be independent of $k$. Correspondingly, for minimum
value of the free energy  one can has
\bea
\label{67}
\frac{F}{L}=(1+B-\mu)-T\sum_{m=1}^{\infty}\int d \lambda f_{m}(\lambda)
\log{(1+ e^{\frac{-\epsilon^{m}(\lambda)}{T}})}=\nn\\
=(1-2\mu)-\epsilon_{\Lambda}(\Lambda=0)
\ena
in correspondence with \cite{S1,W}

In order to analyse the ground state further one should take the 
limit $T\rightarrow 0$ in
equations \re{65}-\re{66} and then put the solution  obtained
 into the expression for the energy \re{57}. If we suppose that
$\epsilon^{n}>0$ for $n>1$ (which can be checked after all by analysing the
corresponding equations for the $\epsilon^{n}$'s) in the sum of 
the integral equations
\re{65}-\re{66}, only one term will contribute in the limit $T\rightarrow0$, 
and we  have
\bea
\label{68}
\epsilon^{1,k}(\lambda)=\epsilon_{0}^{1,k}+\int d \Lambda 
f_{1}(\lambda-\Lambda) \bar \epsilon_{\Lambda}^{k}(\Lambda),\nn\\
\epsilon_{\Lambda}^{k}(\Lambda)=-(\mu_{k}+B)-\int d \lambda 
f_{1}(\Lambda-\lambda) \bar \epsilon^{1,k}(\lambda),
\ena
where
\be
\label{69}
\bar \epsilon_{\Lambda}^{k}(\Lambda)=\left\{\begin{array}{l}
\epsilon_{\Lambda}^{k}(\Lambda),\;\;\;\;\;if\;\;\epsilon_{\Lambda}^{k}
(\Lambda)<0\nn\\
0\;\;\;\;\;\;\;\;\;\;\;\;\;if\;\;\epsilon_{\Lambda}^{k}(\Lambda)>0
\end{array}\right.,\;\;\;
\bar \epsilon^{1,k}(\lambda)=\left\{\begin{array}{l}
\epsilon^{1,k}(\lambda),\;\;\;\;\;if\;\;\epsilon^{1,k}(\lambda)<0\nn\\
0\;\;\;\;\;\;\;\;\;\;\;\;\;\;\;if\;\;\epsilon_{\Lambda}^{k}(\Lambda)>0
\end{array}\right. .
\ee
Hence, in Sutherland's approach we are lead to the concept of two seas:
one for real(non-string) solutions, another for rapidity $\Lambda$
(spin down electrons).

Suppose that $\epsilon^{1,k}(\lambda)=0$ at some point $\lambda=Q$ and
$\epsilon_{\Lambda}^{k}=0$ at $\Lambda=Q_{\Lambda}$, which are the
Fermi rapidities. Then we can write the equations for the ground state
energy as
\be
\label{70}
\frac{E_{0}}{L}=1-4 \pi \sum_{k=1}^{2j+1}\int_{-Q}^{Q} 
d\lambda f_{1}(\lambda)\rho_{p}^{1,k}(\lambda).
\ee

The equations for the density functions also simplify
\be
\label{71}
\sum_{k=1}^{2j+1}\rho_{p}^{1,k}(\lambda)=f_{1}(\lambda)-
\sum_{k=1}^{2j+1}\left(
\int_{-\infty}^{-Q_{\Lambda}}+\int_{Q_{\Lambda}}^{\infty}\right)
d \Lambda f_{1}(\lambda-\Lambda)\sigma_{p}^{k}(\Lambda)
\ee
and
\be
\label{72}
\sum_{k=1}^{2j+1}\sigma_{p}(\Lambda)=\sum_{k=1}^{2j+1}\int_{-Q}^{Q} 
d \lambda f_{1}(\Lambda-\lambda)\rho_{p}^{1,k}(\lambda)
\ee
The Fermi boundaries $Q$ and $Q_{\Lambda}$ are determined from the
particle number and magnetisation equations
\bea
\label{73}
\frac{N}{L}=1-\sum_{k=1}^{2j+1}\left[\int_{-Q}^{Q}d \lambda \rho_{p}^{1,k}
(\lambda)+\left(\int_{-\infty}^{-Q_{\Lambda}}+\int_{Q_{\Lambda}}^{\infty}\right)
d \Lambda \sigma_{p}^{k}(\Lambda)\right],\nn\\
\frac{2S_{z}}{L}=1-\sum_{k=1}^{2j+1}\left[\int_{-Q}^{Q}d \lambda \rho_{p}^{1,k}
(\lambda)+\left(\int_{-\infty}^{-Q_{\Lambda}}-\int_{Q_{\Lambda}}^{\infty}\right)
d \Lambda \sigma_{p}^{k}(\Lambda)\right].
\ena
 
Due to a theorem by Lieb and Mattis \cite{LM}, in zero magnetic field
$S_{z}$ should be zero, which means that spin down electrons 
constitute half of all electrons $M=N/2$ and there is not string states
in the ground state.

We see that the equations for the ground state are invariant under
transformation of the index $k$, which means that we can arbitrarily
distribute the colour over the particles. They are as gauge modes
in the vacuum of gauge theories.

In the half filled case $Q_{\Lambda}=0$ and $Q=\infty$ (as easy to see
in Lai's form of equations \cite{S1}) and following \cite{S1,B} one can
obtain $E_{0}=1-2 \log{2}$. 

It is also possible , following \cite{W}, to consider the excitations 
of the model, introduce the so called shift functions and write down
equations for them. After that, following \cite{EK1}, one can find
the $S$-matrix for the excitations of the model. It follows as above, 
that the $S$-matrix of our model is equal to the $S$-matrix
of ordinary $t-J$ model multiplied by unity operators in additional
colour space.
\section{Conclusions}

We have constructed a one dimensional model which is based on the t-J model 
of strongly
correlated electrons, but which has an additional quantum group symmetry, 
ensuring
the degeneration of the states. We use the Bethe Ansatz technique to investigate
this model. The  equations for density functions, written in the thermodynamic
limit , demonstrate that
the additional degrees of freedom of the model
behave as gauge modes. The presence of these modes, in our opinion, gives
rise a possibility to construct a new type of integrable models, if one has
an interaction between two models of this kind. Also, different 
topological properties,  usually appearing in gauge theories, 
could be present in our model, and would be
interesting to investigate. 



\end{document}